\documentclass[prc,aps]{revtex4}  \usepackage{graphicx} 
\begin{document} 
\draft
\title{Investigation of constraints on few-neutron forces in neutron matter 
by empirical information on the neutron
skin of $^{48}$Ca and $^{208}$Pb  
}
\author{            
Francesca Sammarruca      }                                                                                
\affiliation{ Physics Department, University of Idaho, Moscow, ID 83844-0903, U.S.A. 
}
\date{\today} 
\begin{abstract}
The neutron matter equation of state is calculated from two-neutron forces up
to fifth order of the chiral expansion and the order-by-order convergence of
the predictions is investigated. Based on these equations of state, the binding energies      
and the neutron and proton density distributions in   
$^{208}$Pb and $^{48}$Ca are derived, with particular attention to the neutron skins, the 
focal point of this paper. 
Anticipating future experiments which will provide reliable information on the 
weak charge density in nuclei,       
the theoretical uncertainties and the possibility of  
constraining the size of few-neutron forces in neutron matter are discussed.                                       
\end{abstract}
\maketitle

\section{Introduction} 
\label{Intro} 
Chiral effective field theory (EFT) has become established as a model-independent 
approach to construct nuclear two- and many-body forces in a systematic and internally 
consistent manner~\cite{Wei68,Wein79}. 
Nucleon-nucleon (NN) potentials have been developed from next-to-leading order (NLO, second order) to
N$^3$LO (fourth order)~\cite{NLO,EM03,ME11,EGM05,EHM09}, with the latter reproducing NN data at the high precision 
level. More recently, NN chiral potentials at N$^4$LO have become available~\cite{n4lo,EKM15}. 

Consistent application of these potentials in few- and many-body systems requires 
inclusion of all few- and many-nucleon forces which appear at the given order of chiral EFT,                
a task of greater and greater complexity with increasing 
order. In fact, even today, all two-, three-, and four-nucleon forces of order
greater than three have not yet been applied in an $A > $ 3 system, although
several {\it ab initio} calculations of nuclei and nuclear matter       
based on chiral EFT have been reported. A fairly extensive, 
although likely not exhaustive list is given in Refs.~\cite{NCS,cc1,cc2,cc3,SRG,Stro16,ab1,ab2,ab3,ab4,Cor+,ab5,Soma,ab5a,ab6,ab7,ab8,ab9,ab10,CEA}. Predictions in neutron matter from chiral EFT interactions can be found in Ref.~\cite{kth}.

On the other hand, thanks to recent progress in the development of chiral NN forces~\cite{n4lo,EKM15},
internally consistent calculations can be conducted in the many-body system with two-nucleon forces (2NFs) 
up to fifth order. 
Although the predictions thus obtained are incomplete, they can provide valuable information on 
what is missing. 
Observing the order-by-order convergence of such 2NF based calculations, one can pin down the effect
of 3NFs with uncertainty quantification. 
Together with reliable empirical information on the observables under consideration, the size of the missing 3NFs
can be constrained. 

Neutron-rich systems are especially interesting and are 
receiving considerable attention. Neutron-rich nuclei are
intriguing for many reasons, ranging from the mechanism that controls the
formation of the neutron skin to remarkable correlations with the properties of
compact stars. On the other hand, the properties of these nuclei are, in general, poorly constrained.
However,                           
the electroweak program at the Jefferson Laboratory promises to deliver 
accurate measurements of the neutron skin in $^{208}$Pb and, potentially, $^{48}$Ca. 
Note that, for the latter nucleus, {\it ab initio} calculations are now possible~\cite{hag+}.

The arguments stated above motivate the present work.                              
It is the purpose of this paper to explore to which extent one can estimate the size of three-neutron forces in neutron 
matter using empirical constraints.
After a description of the novel aspects of this work (Sec.~\ref{n4})
and a brief review of previously developed formalism (Sec.~\ref{rev}), 
order-by-order results are shown for nuclear properties in 
$^{208}$Pb and $^{48}$Ca (Sec.~\ref{res}).                                                           
For that purpose, microscopic equations of 
state (EoS) of neutron matter with 2NFs up to 5th order of chiral EFT are
first calculated. The theoretical uncertainties arising from diverse sources are discussed and 
available constraints on the skins of 
$^{208}$Pb and $^{48}$Ca are examined to explore the likelihood that future, more stringent
constraints would allow 
to estimate the size of few-neutron forces in neutron matter.       
Section~\ref{Concl} concludes the paper. 
                                                                     
\section{Nuclear properties from two-nucleon forces up to 5th order } 
\label{tables} 

\subsection{The nucleon-nucleon force in neutron matter at N$^4$LO} 
\label{n4} 

\begin{figure}[!t]
\vspace*{1cm}
\includegraphics[width=12.5cm]{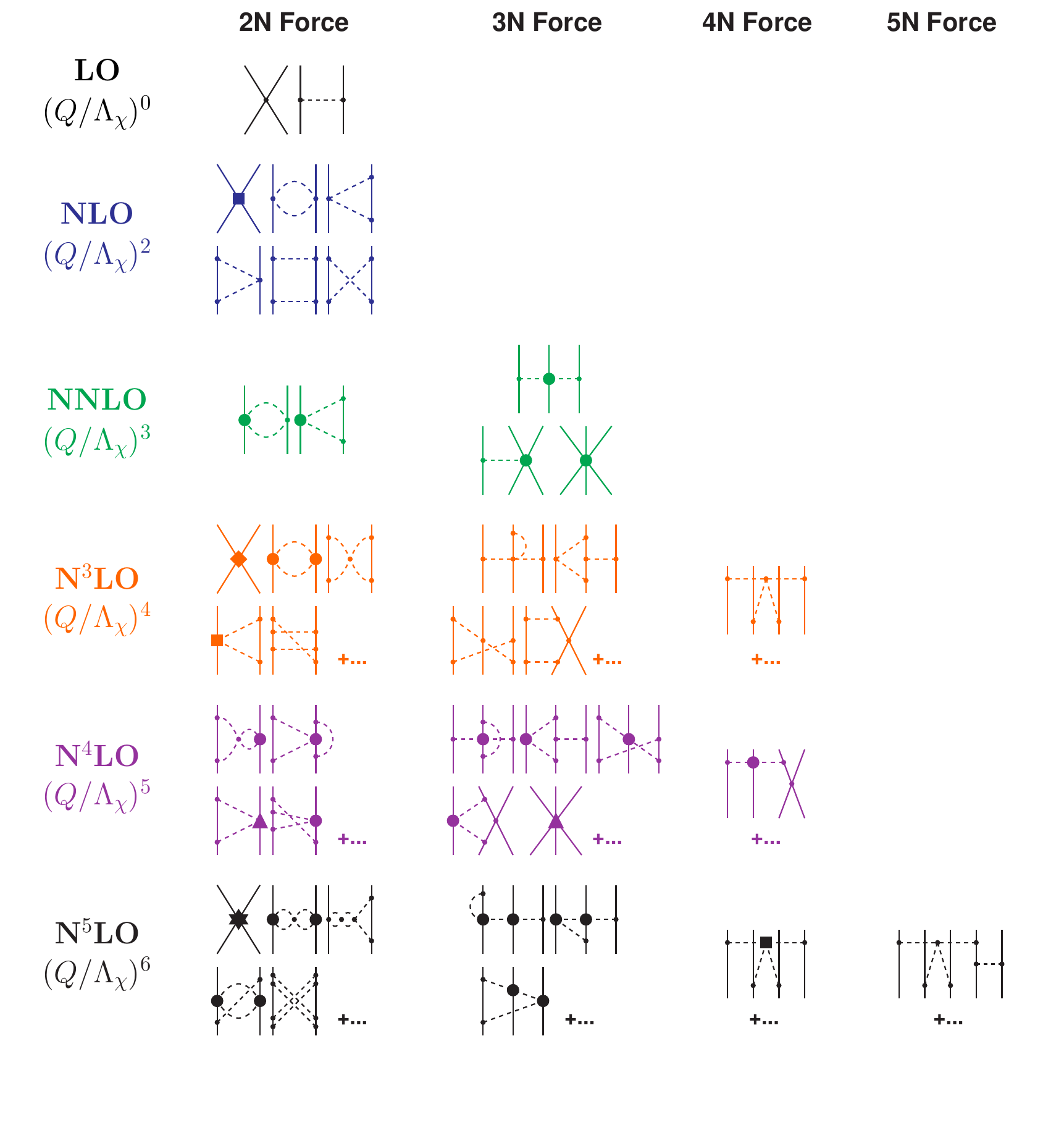}
\vspace*{0.1cm}
\caption{(Color online)                                                                                 
Diagrams of two- and few-nucleon forces appearing at increasing orders of chiral perturbation theory.
} 
\label{diag}
\end{figure}

The neutron matter
EoSs used as input are obtained as in Ref.~\cite{obo} up to fourth order, but without 3NFs. 
An important novel aspect here is the extension of the 2NF to fifth order of chiral EFT.               
The NN interaction employed is the
one at N$^4$LO whose predictions for peripheral partial waves were shown in 
Ref.~\cite{n4lo}. The potential includes one- and two-loop two-pion exchanges and two-loop three-pion 
exchanges as required at this order, see 
Fig.~\ref{diag}.

Although at N$^2$LO the main features of the 
nuclear force can be described reasonably well, it is 
well known that one must go to the next order to achieve high precision.
However, at N$^3$LO (as at 
 N$^2$LO), the chiral 2$\pi$ exchange is still too attractive. It is shown in Ref.~\cite{n4lo} that the 
  2$\pi$ exchange at N$^4$LO is mostly repulsive, which allows an improved description of the $F$ and $G$
partial waves. The overall contribution from the 3$\pi$ exchange is found to be of
 moderate size, suggesting convergence
with regard to the number of exchanged pions. 
The hierarchy of nuclear forces as they emerge at each order of chiral EFT is displayed in Fig.~\ref{diag}. 

\begin{figure}[!t]
\vspace*{1cm}
\includegraphics[width=8.5cm]{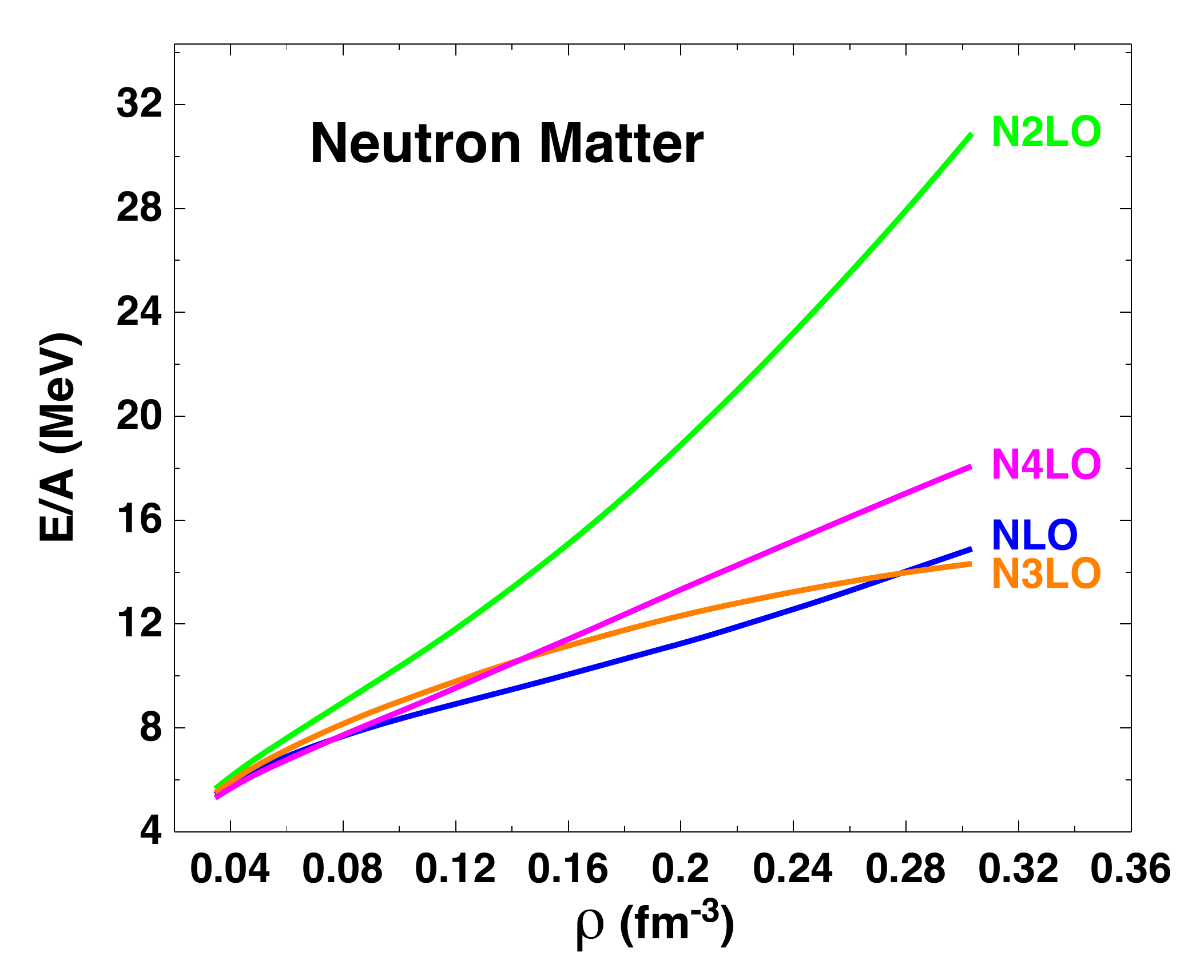}
\vspace*{0.1cm}
\caption{(Color online) Energy per neutron as a function of neutron 
matter density $\rho$ obtained with chiral 2NFs at the indicated 
orders of EFT. 
} 
\label{nm2}
\end{figure}

\begin{figure}[!t]
\vspace*{1cm}
\includegraphics[width=8.5cm]{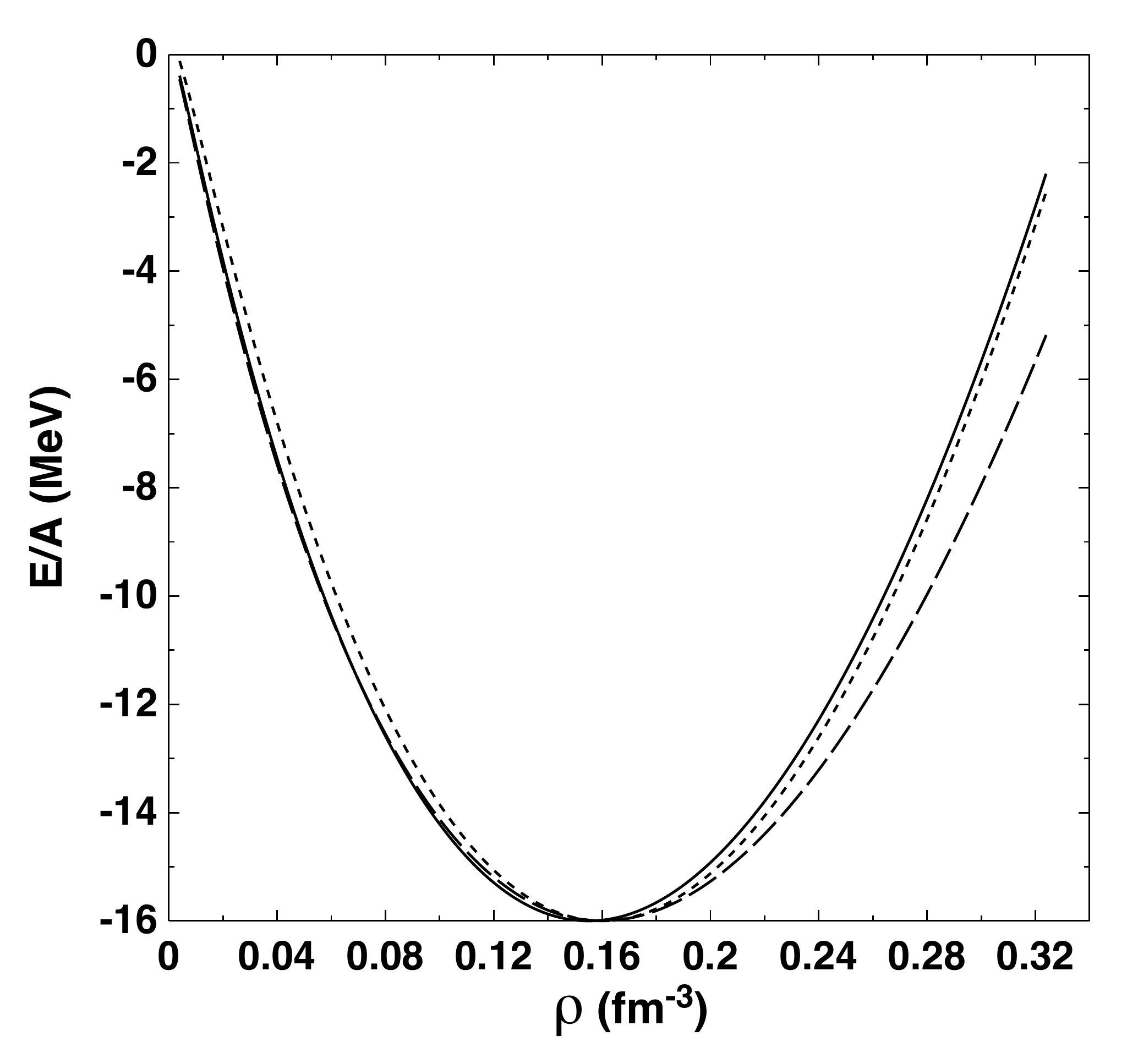}
\vspace*{0.1cm}
\caption{                                                                                 
Phenomenological equations of state for symmetric nuclear matter. Solid: From Ref.~\cite{snm};
Dashed: From Ref.~\cite{Chen}, lower incompressibility; Dotted: 
From Ref.~\cite{Chen}, higher incompressibility. See text for more details. 
} 
\label{eos}
\end{figure}

The neutron matter EoS is calculated within the particle-particle ladder approximation, 
order by order from NLO to 
 N$^4$LO using 2NFs only. The EoS are displayed in Fig.~\ref{nm2}.                     
(Note that                     
the leading order (LO) is not included because it is an extremely poor representation of the nuclear force and thus 
would not add much to the discussion, even in the context of order-by-order convergence.) 
The order-by-order pattern shows a clear signature of convergence: The fifth order correction is 
substantially smaller than the fourth order one. 

\subsection{Brief review of additional tools} 
\label{rev} 
To facilitate the understanding of the results, this section provides a very brief review of previously developed 
tools.      
Nuclear properties are obtained as described in Ref.~\cite{AS03}. Namely, inspired by a liquid 
droplet model, the energy of a          
 nucleus is written in terms of a volume, a surface, and a Coulomb term as 
\begin{equation}
E(Z,A) = \int d^3 r~ e(\rho,\alpha)\rho(r) + 
\int d^3 r f_0|\nabla \rho|^2 +        
\frac{e^2}{4 \pi \epsilon_0}(4 \pi)^2 
\int _0^{\infty} dr' r' \rho_p(r')       
\int _0^{r'} dr r^2 \rho_p(r) \; .  
\label{drop} 
\end{equation} 
In the above equation, 
$\rho$ is the isoscalar density, given by $\rho_n +\rho_p$,     
$\alpha$ is the neutron asymmetry
parameter, $\alpha=\rho_I/\rho$,                                                     
where the isovector density $\rho_I$ is given by $(\rho_n -\rho_p)$.                                    
$e(\rho,\alpha)$ is the energy per particle in 
isospin-asymmetric nuclear matter, written as 
\begin{equation}
e(\rho,\alpha) =                                  
e(\rho,0) + e_{sym}(\rho)\alpha^2 \; ,             
\label{eee}  
\end{equation}
with $e_{sym}(\rho)$ the symmetry energy. 
The density functions for protons and neutrons are obtained by minimizing the value
of the energy, Eq.~(\ref{drop}), with respect to the paramaters of Thomas-Fermi distributions, 
\begin{equation}
 \rho_i(r) = \frac{\rho_0}{1 + e^{(r-a_i)/c_i}} \; , 
\label{TF}  
\end{equation}
with $i=n,p$. The radius and the diffuseness, $a_i$ and $c_i$, respectively, are extracted by minimization 
of the energy while $\rho_0$ is obtained by normalizing the proton(neutron) distribution to $Z$($N$). 
The neutron skin, which is the object of this investigation, is defined as 
\begin{equation}
 S = R_n - R_p \; ,                                 
\label{rnrp}   
\end{equation}
where $R_n$ and $R_p$ are the $r.m.s.$ radii of the neutron and proton density distributions, 
\begin{equation}
 R_i = \Big ( \frac{4 \pi}{T} \int_0 ^{\infty} \rho_i(r) r^4 \; dr \Big )^{1/2} \; , 
\label{rms}   
\end{equation}
where $T$= $N$ or $Z$. 
This method has the advantage of allowing for a         
very direct connection between the EoS and the properties of finite nuclei. It was used in Ref.~\cite{AS03} in 
conjunction with relativistic meson-theoretic potentials and found to yield realistic predictions for binding 
energies and charge radii. 
The constant $f_0$ in the surface term is typically obtained from fits to $\beta$-stable nuclei and found to be about
60-70 MeV fm$^5$~\cite{Oya2010}. How this uncertainty impacts the corresponding predictions will be discussed below. 

The isospin-symmetric part of the EoS in Eq.~(\ref{eee}) is taken from phenomenology~\cite{snm} to maintain the focus on
the microscopic neutron matter predictions.                 
The EoS employed here for symmetric matter was obtained from empirically determined values of 
characteristic constants in homogeneous matter at saturation and subsaturation~\cite{snm}, with isoscalar quantities (and also isovector ones) found to be very well constrained. 
 At the low densities  
probed by the neutron skin, one might expect that equations of state constructed to reproduce closely empirical properties 
 will not be appreciably different from one another, as is confirmed in Fig.~\ref{eos}.
Nevertheless, to estimate the uncertainty associated with different phenomenological parametrizations of
the symmetric matter EoS,                                                                   
the phenomenological EoS from Ref.~\cite{Chen}, designed to describe                
 both isospin symmetric and asymmetric nuclear matter, will be used in 
addition. For symmetric nuclear matter, it is given as 
\begin{equation}
e(\rho,0) = \frac{3 \hbar^2}{10 m} \Big (\frac{3 \pi^2}{2}\rho \Big )^{2/3} + \frac{\alpha}{2} 
\frac{\rho}{\rho_0} + \frac{\beta}{\sigma + 1} \Big (\frac{\rho}{\rho_0} \Big )^{\sigma} \; , 
\label{snm2}   
\end{equation}
where $\rho_0$ = 0.16 fm$^{-3}$, the energy at saturation is very close to -16 MeV, and 
$\alpha$, $\beta$, and $\sigma$ are expressed in terms of the incompressibility 
$K_0$~\cite{Chen} whose commonly accepted value is 240 $\pm$ 20 MeV. 
In Fig.~\ref{eos}, we show the equation of state from Ref.~\cite{snm} (solid), whose parameters are fitted to the 
central values of the constraints (e.~g. $K_0$=240 MeV), in comparison to the one from Ref.~\cite{Chen}
with parameters corresponding to $K_0$=(240-20) MeV (dashed) or 
$K_0$=(240+20) MeV (dotted).  The predictions shown by the dotted curve, which 
appear to differ more noticeably from the solid curve at subsaturation to saturation densities, will be used here to
estimate the uncertainty arising from the choice of the phenomenological EoS. 
In fact, several tests confirmed that the larger differences between the solid and the dashed 
curves at suprasaturation densities are essentially insignificant for the neutron skin investigations performed
here. 
Note, further, that I am not considering a family of {\it theoretical} EoS for symmetric matter, since I wish to 
keep out of this investigation any model dependence which may arise from those. 

\section{Results} 
\label{res} 

\begin{figure}[!t]
\vspace*{1cm}
\includegraphics[width=7.5cm]{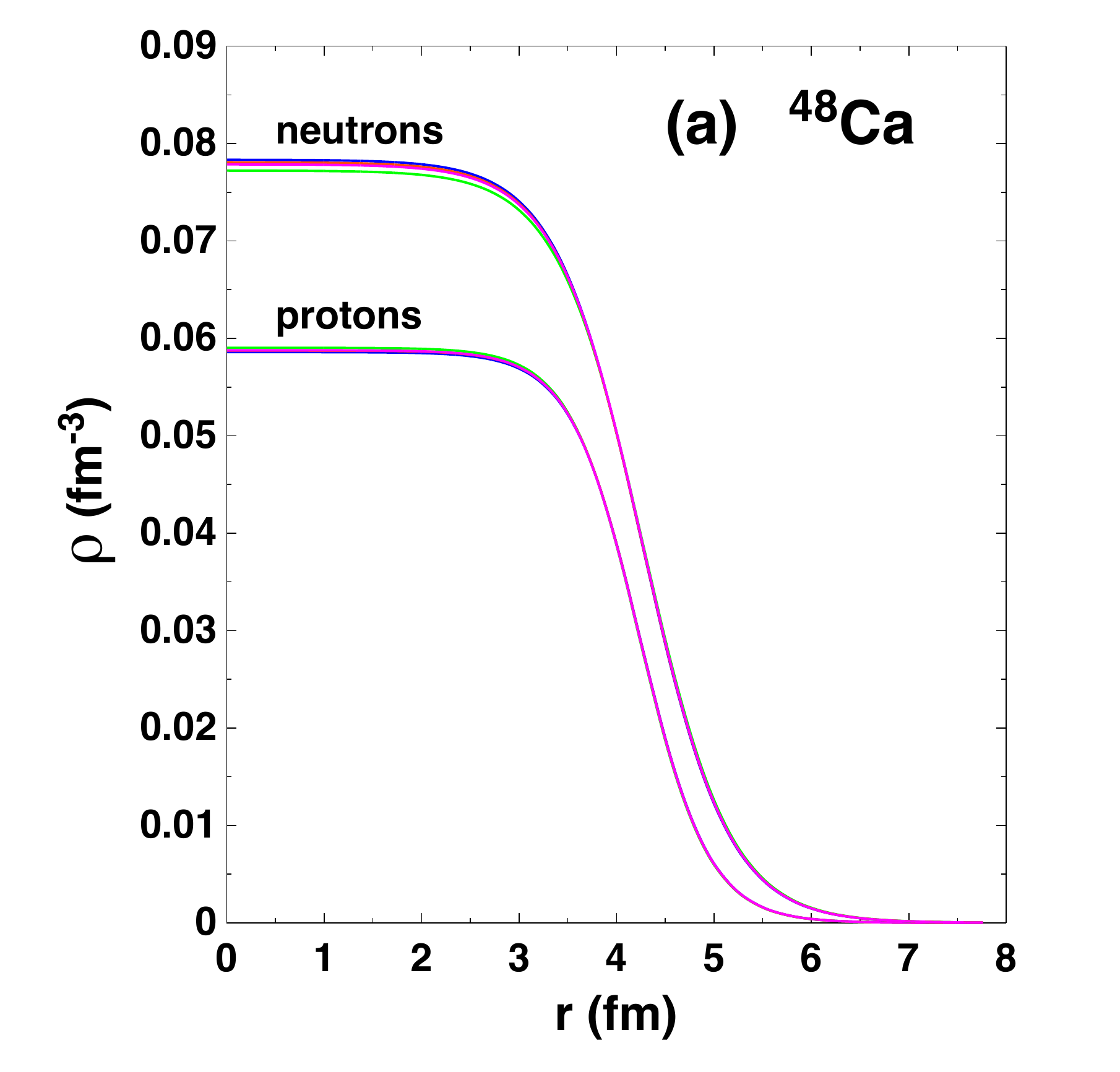}
\includegraphics[width=7.5cm]{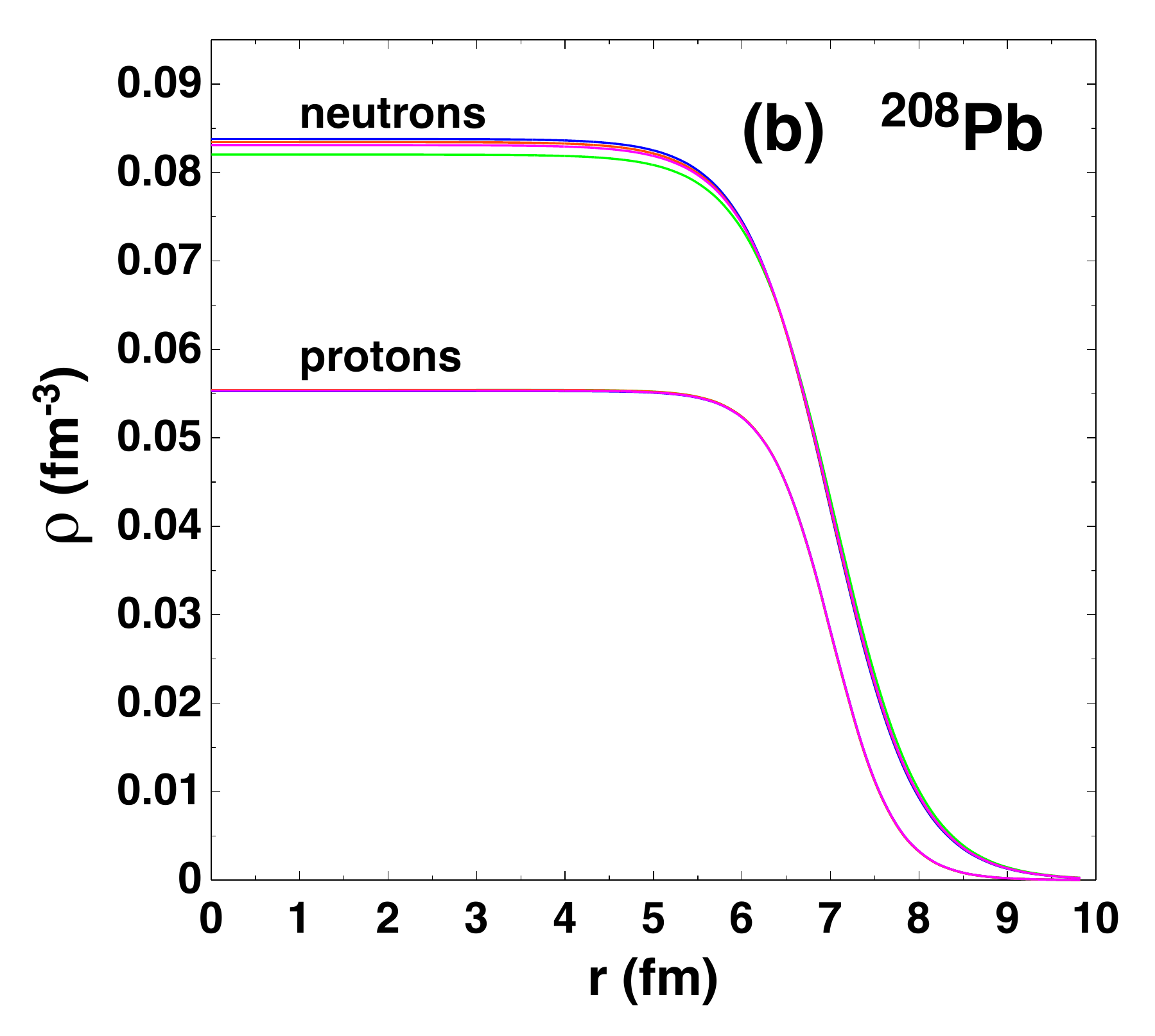}
\vspace*{0.1cm}
\caption{(Color online) Density distributions for neutrons and protons 
in {\bf (a)} $^{48}$Ca and {\bf (b)}                                                         
 $^{208}$Pb, for different orders of chiral EFT from NLO to N$^4$LO. Color code as in 
Fig.~\ref{nm2}. For protons, the various curves cannot be distinguished on the scale of the figures. For neutrons,
the lowest (green) curve represents the N$^2$LO result. 
}  
\label{rho}
\end{figure}

As a first look into order-by-order convergence, I begin this section by showing, 
cf.~Fig.~\ref{rho}, the two-parameter Fermi functions obtained for neutron and proton densities from 
NLO to N$^4$LO. 
Obviously, order-by-order differences cannot be discerned on the scale of the figure, with the exception 
of the neutron densities.                                                    
The green curve (lower curve) reflects the stronger repulsion (hence, lower central densities) of the
EoS at N$^2$LO, cf.~Fig.~\ref{nm2}. 
The predictions in Fig.~\ref{rho} are obtained
with $\Lambda$=450 MeV, but order-by-order differences remain small when varying the scale, as will be discussed
below.

Next, I will focus on 
the binding energy per nucleon, the charge radius, the
point proton  
and neutron radii, and the neutron skin for $^{48}$Ca and
$^{208}$Pb. 
I will consider truncation error, sensitivity to cutoff variations, as well as uncertainties associated with
the density functional including the choice of the phenomenological EoS.                

With regard to the cutoff parameter, which appears in the 
regulator function             
\begin{equation}
f(p',p) = exp[-(p'/\Lambda)^{2n} - 
(p/\Lambda)^{2n}] \;, 
\label{cut} 
\end{equation}
values between 450 and about 600 MeV                         
will be considered, with 550 MeV being the largest available at 
 N$^4$LO.                                                                             
 Note that these values are below the breakdown scale of 
chiral EFT~\cite{dpot}. 
Other analytical expressions are possible for the regulator function~\cite{dpot}. It will be interesting  
to include these potentials when they become available to the community at large. 
Furthermore, 
coordinate-space potentials have been developed up to N$^2$LO for the 2$\pi$-exchange contributions and 
up to N$^3$LO for the contact terms~\cite{Pia15}. Therefore, a consistent study at 
N$^3$LO and beyond, as the one undertaken here, would not be possible with the potentials from Ref.~\cite{Pia15}.

 The results for $^{48}$Ca, using the $f_0$ parameter [cf.~Eq.~(\ref{drop})] on the lower side, are shown 
in Table~\ref{tab1}. 
One can see that these nuclear properties show         
good convergence tendency at N$^4$LO.
Similar comments apply to Table~\ref{tab2}, where the predictions differ from those in Table~\ref{tab1} only in 
the larger value of $f_0$, which introduces more repulsion in the liquid droplet model. Binding energy values are
smaller by  a few percent and the r.m.s. radii remain very close to those in Table~\ref{tab1}. Once 
again, all properties show a clear signature of convergence towards N$^4$LO.
The results for 
$^{208}$Pb, which are given in                                               
Table~\ref{tab3} and Table~\ref{tab4}, show trends very similar to those observed in 
$^{48}$Ca.                                                                 

In the uncertainty analysis which follows, 
the results at N$^3$LO will be taken as the ``final" predictions,
since the truncation error at this order can be reliably estimated from the knowledge of the predictions at N$^4$LO.
For $^{48}$Ca at N$^3$LO,                                                      
using the smaller value of $f_0$ (cf.~Table~\ref{tab1}) and averaging the results for the different cutoffs yields 
$S$ = 0.148 $^{+0.002}_{-0.003}$ fm, whereas a similar average 
for the larger value of $f_0$ (cf.~Table~\ref{tab2}) gives 
$S$ = 0.162 $^{+0.003}_{-0.003}$ fm.                                      

At N$^3$LO, the truncation error is given by the difference between
the predictions at N$^3$LO and those at                                                         
N$^4$LO.                                                                     
 Or, in other words: The N$^4$LO correction is the N$^3$LO uncertainty. 
Applying this reasoning, next I take                                                                   
the difference between cutoff-averaged predictions at 
N$^3$LO and N$^4$LO, respectively, and determine the truncation error to be about 0.001 fm, showing 
 that the results are very well converged with respect to the chiral expansion of the two-nucleon force. 

Further, to account for the 
uncertainty arising from the parameter $f_0$ in the droplet model,      
the central values given above are averaged, which yields ${\bar S}$= 0.155 $\pm$ 0.007 fm.
                          
The same steps are then repeated using 
another phenomelological EoS for symmetric matter (see discussion above, at the end of 
Sect.~\ref{rev}). Using 
the smaller value of $f_0$ and averaging the results for the different cutoffs,       
I obtain $S$ = 0.155 $^{+0.002}_{-0.003}$ fm, whereas a similar average 
for the larger value of $f_0$ gives 
$S$ = 0.169 $^{+0.0024}_{-0.003}$ fm. The truncation error is again very small (about 0.001 fm).
Averaging the central values as before  
yields ${\bar S}$= 0.162 $\pm$ 0.007 fm.

Finally, 
combining the results obtained with two phenomelogical EoS and calculating the total error in 
quadrature, 
the prediction based on chiral 2NFs at N$^3$LO is found to be: 
\begin{equation}
{\bar S}_{2NF}(^{48}Ca) = (0.159 \pm 0.009) \; \mbox{fm} \; .                                                  
\label{ca_th} 
\end{equation}

An identical analysis for 
$^{208}$Pb yields                                                             
\begin{equation}
{\bar S}_{2NF}(^{208}Pb) =                                                                  
 (0.14 \pm 0.01) \; \mbox{fm} \; .                                                   
\label{pb_th} 
\end{equation}

\begin{table}                
\centering
\begin{tabular}{|c||c||c|c|c|c|c|}
\hline
 Order & Cutoff(MeV) & $B/A$(MeV) & $r_{ch}$(fm) & $r_p$(fm) & $r_n$ (fm) & $S$ (fm) \\ 
\hline     
\hline
 NLO & 450& 8.735  & 3.620    & 3.517  & 3.655   & 0.138  \\ 
 NLO & 500 & 8.734  & 3.620    & 3.517  & 3.656   & 0.138  \\ 
 NLO & 600 & 8.735  & 3.621    & 3.518  & 3.658   & 0.140  \\ 
\hline
N$^2$LO  & 450 & 8.693  & 3.613    & 3.510  & 3.672   & 0.162  \\ 
N$^2$LO & 500&  8.690  & 3.613    & 3.510  & 3.674   & 0.164  \\ 
N$^2$LO & 600& 8.686  & 3.612    & 3.509  & 3.675   & 0.166  \\ 
\hline
N$^3$LO & 450&  8.723  & 3.618    & 3.515  & 3.660    &0.145   \\ 
N$^3$LO & 500& 8.715  & 3.617    & 3.514  & 3.663    &0.149   \\ 
N$^3$LO & 600& 8.714  & 3.616    & 3.513  & 3.663    &0.150   \\ 
\hline
N$^4$LO &450  & 8.728  & 3.618    & 3.515 & 3.661   & 0.146  \\ 
N$^4$LO &500 & 8.724  & 3.617    & 3.514 & 3.661   & 0.147  \\ 
N$^4$LO &550 & 8.722  & 3.617    &3.514  & 3.662.  & 0.148  \\ 
\hline
\end{tabular}
\caption                                                    
{ Binding energy per nucleon ($B/A$), charge radius ($r_{ch}$), proton and neutron 
 point radii ($r_p$ and $r_n$, respectively), and neutron skin ($S$) 
of $^{48}$Ca. The predictions are obtained from a microscopic neutron matter equation of state including
only two-neutron forces at the specified orders of chiral effective field theory. The value of $f_0$ in 
Eq.~(\ref{drop}) is 60 MeV fm$^5$. 
} 
\label{tab1}
\end{table}

\begin{table}                
\centering
\begin{tabular}{|c||c||c|c|c|c|c|}
\hline
Cutoff (MeV) & Order & $B/A$(MeV) & $r_{ch}$(fm) & $r_p$(fm) & $r_n$ (fm) & $S$ (fm) \\ 
\hline     
\hline
NLO &450 & 8.362  & 3.659    & 3.557  & 3.708   & 0.152  \\ 
NLO &500 & 8.362  & 3.659    & 3.557  & 3.709   & 0.152  \\ 
NLO &600 & 8.363  & 3.659    & 3.557  & 3.711   & 0.154  \\ 
\hline
N$^2$LO &450 & 8.324  & 3.651    & 3.549  & 3.725   & 0.176  \\ 
N$^2$LO &500 & 8.321  & 3.651    & 3.549  & 3.727   & 0.178  \\ 
N$^2$LO &600 & 8.318  & 3.650    & 3.548  & 3.728   & 0.180  \\ 
\hline
N$^3$LO &450 & 8.351  & 3.656    & 3.554 & 3.713    &0.159   \\ 
N$^3$LO &500 & 8.344  & 3.655    & 3.553 & 3.716    &0.163   \\ 
N$^3$LO &600 & 8.343  & 3.654    & 3.552 & 3.717    &0.165   \\ 
\hline
N$^4$LO &450 & 8.356  & 3.656    & 3.554  & 3.714   & 0.160  \\ 
N$^4$LO &500 & 8.353  & 3.655    & 3.554  & 3.714   & 0.161  \\ 
N$^4$LO &550 & 8.350  & 3.655    & 3.553  & 3.715   & 0.162  \\ 
\hline
\end{tabular}
\caption                                                    
{Same as Table~\ref{tab1}, but using $f_0$=                                                
70 MeV fm$^5$.                                                                            
} 
\label{tab2}
\end{table}

\begin{table}                
\centering
\begin{tabular}{|c||c||c|c|c|c|c|}
\hline
Cutoff(MeV) & Order & $B/A$(MeV) & $r_{ch}$(fm) & $r_p$(fm) & $r_n$ (fm) & $S$ (fm) \\ 
\hline     
\hline
 NLO &450 & 7.966  & 5.645    & 5.580  & 5.690   & 0.110  \\ 
NLO &500 & 7.963  & 5.646    & 5.581  & 5.693   & 0.112  \\ 
NLO &600 & 7.960  & 5.649    & 5.584  & 5.701   & 0.117  \\ 
\hline
N$^2$LO &450 & 7.862  & 5.643    & 5.577  & 5.730   & 0.154  \\ 
N$^2$LO &500 & 7.853  & 5.643    & 5.578  & 5.735   & 0.158  \\ 
N$^2$LO &600 & 7.844  & 5.644    & 5.578  & 5.740   & 0.162  \\ 
\hline
N$^3$LO &450 & 7.936  & 5.642    & 5.576  & 5.699    &0.123   \\ 
N$^3$LO &500 & 7.917  & 5.643    & 5.577  & 5.708    &0.130   \\ 
N$^3$LO &600 & 7.914  & 5.641    & 5.575  & 5.708    &0.132   \\ 
\hline
N$^4$LO &450 & 7.943  & 5.645    & 5.579   & 5.705    & 0.125  \\ 
N$^4$LO &500 & 7.937  & 5.643    & 5.577   &5.703     & 0.126  \\ 
N$^4$LO &550 &7.930   & 5.642    & 5.576   &5.705     & 0.129  \\ 
\hline
\end{tabular}
\caption                                                    
{ Same as Table~\ref{tab1}, but for $^{208}$Pb.                                            
} 
\label{tab3}
\end{table}

\begin{table}                
\centering
\begin{tabular}{|c||c||c|c|c|c|c|}
\hline
Cutoff(MeV) & Order & $B/A$(MeV) & $r_{ch}$(fm) & $r_p$(fm) & $r_n$ (fm) & $S$ (fm) \\ 
\hline     
\hline
NLO& 450 & 7.741  & 5.671    & 5.605  & 5.729   & 0.124  \\ 
NLO &500 & 7.739  & 5.672    & 5.606  & 5.732   & 0.125  \\ 
NLO &600 & 7.736  & 5.675    & 5.609  & 5.740   & 0.130  \\ 
\hline
N$^2$LO &450 & 7.642  & 5.667    & 5.602  & 5.770   & 0.169  \\ 
N$^2$LO &500 & 7.634  & 5.667    & 5.602  & 5.775   & 0.173  \\ 
N$^2$LO &600 & 7.625  & 5.668    & 5.603  & 5.780   & 0.177  \\ 
\hline
N$^3$LO &450 & 7.712  & 5.667    & 5.602  & 5.738    &0.137   \\ 
N$^3$LO &500 & 7.694  & 5.668    & 5.602  & 5.747    &0.145   \\ 
N$^3$LO &600 & 7.692  & 5.666    & 5.601  & 5.747    &0.147   \\ 
\hline
N$^4$LO &450 & 7.719  & 5.670    & 5.605  & 5.744   & 0.139  \\ 
N$^4$LO &500 & 7.714  & 5.668    & 5.603  & 5.743   & 0.140  \\ 
N$^4$LO &550 & 7.707  & 5.667    & 5.602  & 5.744   &0.143   \\ 
\hline
\end{tabular}
\caption                                                    
{ Same as Table~\ref{tab2}, but for $^{208}$Pb.                                            
} 
\label{tab4}
\end{table}

The question to be 
addressed next is whether one can constrain
the effect from few-neutron forces in neutron matter using                         
these well-converged results based on 2NFs only and empirical 
information. 
Some comments are in place here concerning the nature of the contributions one may 
potentially constrain. In principle, four- and higher-body forces are included in the missing terms.          
However, it is reasonable to expect that by far the largest contribution would be from 3NFs. In fact, 
chiral perturbation theory offers a justification for why higher-body forces should be smaller, 
since they appear at higher order in the expansion. 
An investigation aimed at constraining 3NFs exploiting chiral 2NFs can be found in Ref.~\cite{3a}.

Accurate measurements of the neutron skin of 
 $^{208}$Pb from PREX experiments~\cite{Jlab} (and, potentially, C-REX experiments for
 $^{48}$Ca~\cite{CREX}), are expected but               
 not yet available. Thus, I will start from current information and project a near-future scenario
when accurate measurements of neutron radii become available from parity-violating electron scattering experiments.

Table~\ref{tab6} displays some representative empirical results for the neutron skin thickness of 
 $^{48}$Ca and                                         
 $^{208}$Pb extracted by a variety of methods (see corresponding citations). 
Reference~\cite{Fried2012} makes use of pionic probes and total $\pi ^+$ reaction cross sections between
0.7 and 2 GeV/c. The first two values for calcium displayed in Table~\ref{tab6} are obtained with pionic atoms 
adopting two different versions of the neutron density~\cite{Trz+}. 
The last entry for calcium was obtained from analyses of $\pi^+$ and $\pi^-$ scattering across the 
(3,3) resonance~\cite{Gibbs}. 
The same comment applies to the first two
table entries for lead. The authors of Ref.~\cite{GNO} also make use of pionic atom potentials while varying 
radial parameters of the neutron distibutions. The third entry for lead in Table~\ref{tab6} is a weighted
average of their analysis as well as results from previous models. 
 The last $^{208}$Pb entry is extracted from symmetry energy constraints and 
is consistent with a broad set of skin measurements based on antiprotonic atoms, Pigmy Dipole resonances,
electric dipole polarizability, and proton elastic                    
scattering~\cite{Tsang+}. 

By averaging the values for 
$^{48}$Ca of Table~\ref{tab6} and calculating the error in quadrature, one can estimate the current 
knowledge of the neutron skin in $^{48}$Ca as:   
\begin{equation}
{\bar S}_{emp}(^{48}Ca) = (0.13  \pm 0.03) \; \mbox{fm} \; .
\label{ca_exp}
\end{equation}
The difference between theory and experiment then comes out to be:
\begin{equation}
|{\bar S}_{2NF}(^{48}Ca) -                                  
{\bar S}_{emp}(^{48}Ca)| = (0.03 \pm 0.03) \; \mbox{fm} \; .
\end{equation}
Obviously, the difference between the central values from Eqs.~(\ref{ca_exp}) and (\ref{ca_th}) is about the same as 
the uncertainty and, therefore, current
empirical determinations of the neutron skin of 
$^{48}$Ca cannot pin down the effect of the 3NF on neutron matter.                   

The situation is similar for 
 $^{208}$Pb, where the average of the empirical values shown in Table~\ref{tab6} results in 
\begin{equation}
{\bar S}_{emp}(^{208}Pb) = (0.16  \pm 0.04) \; \mbox{fm} \; .
\label{pb_exp}
\end{equation}
Here, the difference between theory and experiment is:              
\begin{equation}
|{\bar S}_{2NF}(^{208}Pb) -                                  
{\bar S}_{emp}(^{208}Pb)| = (0.02 \pm 0.04) \; \mbox{fm} \; ,
\end{equation}
which is smaller than the uncertainty.                                                  

To summarize, from the present analysis 
one may conclude that a measurement of the neutron skin can provide a constraint for the effect of
3NFs on neutron matter if the experimental uncertainty is 
$\Delta {\bar S}_{emp} < |{\bar S}_{2NF} - {\bar S}_{emp}|$. 
Based on the above values, one may conclude that future experiments on the neutron skin of 
 $^{208}$Pb                                                                                   
 (or $^{48}$Ca) should aim for an uncertainty                                             
$\Delta {\bar S}_{emp} < $ 0.03 fm to provide a useful constraint on 3NFs in neutron matter.

\begin{table}                
\centering
\begin{tabular}{|c||c|c|}
\hline
Nucleus & $S$ (fm) & Source    \\ 
\hline     
\hline
$^{48}$Ca & 0.13 $\pm$ 0.06  & Ref.~\cite{Fried2012} \\ 
          & 0.16 $\pm$ 0.07  & Ref.~\cite{Fried2012} \\ 
          & 0.11 $\pm$ 0.04  & Ref.~\cite{Gibbs} \\ 
\hline
$^{208}$Pb & 0.15 $\pm$ 0.08 & Ref.~\cite{Fried2012} \\ 
           & 0.14 $\pm$ 0.10 & Ref.~\cite{Fried2012} \\ 
           & 0.18 $\pm$ 0.05 & Ref.~\cite{GNO} \\ 
           & 0.18 $\pm$ 0.05  & Ref.~\cite{Tsang+} \\ 
\hline
\end{tabular}
\caption                                                    
{Empirical values for the neutron skin of                                      
 $^{48}$Ca and                                         
 $^{208}$Pb taken from various sources.                    
} 
\label{tab6}
\end{table}
\section{Summary and Conclusions}                                                                  
\label{Concl} 

Predictions which cannot be stated with appropriate theoretical uncertainty are no longer consistent
with contemporary standards. 
With chiral EFT, one can reliably estimate the truncation error at each order of the chiral expansion.     
Being able to do so is crucial to guide future measurements.

In this work, 
the neutron matter EoS applying chiral 2NFs up to fifth order has been calculated, thus extending previous predictions.
Using as input the microscopic neutron matter EoS from second order to fifth order,                  
 the order-by-order convergence pattern of the neutron skin in 
 $^{48}$Ca and                                         
 $^{208}$Pb was explored.                                                
It turns out that the uncertainty with regard to the chiral expansion of the 2NF up to N$^4$LO is 0.001 fm 
for both nuclei, which reflects an excellent degree of convergence concerning the Hamiltonian. Including 
(non-local) cutoff variations and the 
error from the many-body method applied, the overall uncertainty of these predictions comes out to be
about 0.01 fm for 
 $^{48}$Ca and                                         
 $^{208}$Pb. This small theoretical uncertainty of the 2NF based predictions should, in principle, allow
to pin down the effects of the missing 3NFs, if the empirical determination carries a sufficiently small error, 
which is not the case with present constraints.  
 This analysis finds that the experimental error of 
neutron skin determinations for these nuclei should be less than 0.03 fm to be effective in constraining       
missing contributions from few-nucleon forces. 
These findings can be a useful guideline
for planners of future PREX/CREX experiments.                                           

Before closing, it is important to remind the reader that this analysis                                         
will be broadened in the near future. In particular:
\begin{itemize}
\item 
The uncertainty analysis 
should be extended 
to include a full variation of the regulator function, namely both scheme and scale; 
\item 
 At this time, the EoS of
symmetric matter has been kept fixed to an empirical one, in order to maintain the focus on the possbility to constrain three-neutron forces in neutron matter. The information obtained in the present study will be useful when moving on to a similar investigation which employs, instead, fully microscpic EoS of symmetric matter at each chiral order.              
\end{itemize}

\section*{Acknowledgments}
This work was supported by 
the U.S. Department of Energy, Office of Science, Office of Basic Energy Sciences, under Award Number DE-FG02-03ER41270.

\end{document}